\documentclass[runningheads]{llncs}
\usepackage[T1]{fontenc}
\usepackage{amsfonts}
\usepackage{amsmath}
\usepackage{graphicx}
\usepackage{algorithm}
\usepackage{ wasysym }
\usepackage{algpseudocode}
\newcommand{\Endproof}{\hfill$\Box$\\}
\newcommand{\Beginproof}{{\em Proof.}  }

\begin{document}
\title{Quantum Hashing Circuit Optimization for Arbitrary Qubit Connectivity Graphs Based on 1-Covering Path\thanks{The research has been supported by Russian Science Foundation Grant 25-11-00366, \url{https://rscf.ru/project/25-11-00366/}
The research in Section 4.1 by M.Z. has been partially supported by the Italian Ministero delle Imprese e del Made in Italy (MIMIT) under the project SMART•E - piattaforma per l’IoT Maintenance, il Facility e l’Asset Management dell’industria 4.0, grant number FTE0000382 (CUP: B47H22004430008, COR: 22573728).}}
%\title{Implementation of a Quantum Automaton Based on Quantum Fingerprinting for a Real Device with an Arbitrary Qubits Connection Graph}
\author{Kamil Khadiev\inst{1,2}\orcidID{0000-0002-5151-9908} \and
Aliya Khadieva\inst{1,2}\orcidID{0000-0003-4125-2151} \and
Vadim Sagitov\inst{1}%\orcidID{2222--3333-4444-5555}
\and
Kamil Khasanov\inst{1}
\and
Mansur Ziatdinov\inst{3}\orcidID{0000-0001-7415-2726}
}
\authorrunning{K. Khadiev et al.}
\institute{Kazan Federal University, Kazan, Tatarstan, Russia \and
Zavoisky Physical-Technical Institute,
FRC Kazan Scientific Center of RAS, Kazan, Russia\and
University of Messina, Messina, Italy\\
\email{kamilhadi@gmail.com}}
\maketitle              % typeset the header of the contribution
\begin{abstract}
One of the obstacles to the widespread adoption of quantum computing is the problem of efficient circuit synthesis. Current quantum hardware has limited connections between qubits, with each qubit connected to only a few others. This means that the circuit has to be transformed to accommodate this. In this paper, we present an algorithm that converts a circuit containing a sequence of CNOT gates into a form that is suitable for arbitrary quantum computer architectures. 
Although we demonstrate the algorithm only in the context of quantum fingerprinting, similar gate sequences are prevalent in quantum algorithms; for instance, they are present in the textbook quantum Fourier transform. 
We present a quantum circuit implementation of the quantum hashing algorithm (quantum fingerprinting algorithm) for a quantum device with restrictions on the application of two-qubit gates that are expressed as a qubit connectivity graph. As an example of usage of the technique, we apply it to  quantum finite automata recognizing the unary $MOD_p=\{a^\ell: \ell \bmod p=0\}$ language, and the $EQ_p=\{a^\ell b^r: \ell \equiv r \pmod p\}$ language.
Given the enhancements that our algorithm provides~-- for instance, in one case it achieves a 16\%--17\% decrease in CNOT circuit cost~-- we believe it could also be useful in a broader quantum compilation context.

%We present a quantum circuit implementation of the quantum hashing algorithm (quantum fingerprinting algorithm) for a quantum device with restrictions on the application of two-qubit gates that are presented by a qubit connectivity graph. As an example of usage of the technique, we apply it to  quantum finite automata recognizing the unary $MOD_p=\{a^\ell: \ell $ mod $p=0\}$ language, and the unary $EQ_p=\{a^\ell b^r: \ell=r ($ mod $p)\}$ language. These are simple examples, but they show all the main properties of the quantum hashing algorithm. We present an optimization technique for shallow circuit for quantum hashing, and consider an arbitrary connected graph for qubits.

%123 We present a quantum circuit implementation of a quantum finite automaton recognizing the unary $MOD_p=\{a^\ell: \ell $ mod $p=0\}$ language. We construct a circuit for a quantum device with restrictions on the application of two-qubit gates that is specified by a qubit connectivity graph. The presented implementation can be used for any algorithm based on quantum fingerprinting or quantum hashing. We present an optimization technique for standard and shallow circuits for quantum hashing, and consider an arbitrary connected graph on qubits.
\end{abstract}
%
%
%
%%%%%%%%%%%%%%%%%%%%%%%%%%%%%%%%%%%%%%%%%%%%
%              Introduction               %
%%%%%%%%%%%%%%%%%%%%%%%%%%%%%%%%%%%%%%%%%%%%
\section{Introduction}
\emph{Quantum computing} \cite{nc2010,a2017,aazksw2019part1} is one of the hot topics in computer science of the last decades.
%
%One of the techniques that allows us to develop efficient size quantum automata is quantum fingerprinting or quantum hashing. 
%
One of the techniques that allows us to develop space-efficient quantum algorithms is quantum fingerprinting or quantum hashing \cite{akvz2025}. 
This technique is well known, and it allows us to compute a short hash or fingerprint that identifies a (potentially long) string of data with high probability. The nature of the technique and its implementation is similar to the QFT algorithm \cite{av2020,k2024aliya}. 
The probabilistic (randomized) technique was developed by Freivalds \cite{Fre79}. Then,  Ambainis and Freivalds \cite{af98} developed its quantum counterpart for automata that was improved by Ambainis and Nahimovs in \cite{an2008,an2009}. Later, Buhrman et al.~in \cite{bcwd2001} provided an explicit definition of quantum fingerprinting for constructing an efficient quantum communication protocol for equality testing. The technique was applied for branching programs (non-uniform automata-like model) 
\cite{agkmp2005,ag05,av2009}. 
%\cite{agkmp2005,ag05,av2009,av2011,av2013}. 
Later, they developed the concept of cryptographic quantum hashing
\cite{av2013hash,aavz2016,akvz2025}.
%\cite{av2013hash,av2014,aav2014,aa2015,aav2016,aav2018,aav2020,aavz2016,v2016binary,vlz2017,aakv2018,vvl2019}.R
%Then, different versions of hashing functions were applied by Ablayev, Vasiliev, Ziiatdinov, Zinnatullin and other researchers \cite{v2016,av2020,z2016,z2016group,z2023}. The technique was also extended for qudits \cite{av2022}.
This approach was widely used in different areas like stream processing algorithms \cite{l2009,l2006}, query model algorithms \cite{aaksv2022,asa2024}, online algorithms \cite{kk2019disj,kk2022}, branching programs \cite{kk2017,kkk2022,agky16}, developing quantum devices \cite{v2016model}, automata (discussed earlier, \cite{af98,AY12,AY11A2021,kz2023,gy2017,YS10A,gy2018,gy2015}), machine learning  \cite{zfsd2025,zd2025} and others.

Turaykhanov et al. implemented the technique in a photon-based real quantum device \cite{avsak2022,tavak2021}. The alternative implementation for photon-based real quantum devices was developed by Plachta et al. \cite{phyf2022} and Zhao, et al. \cite{zllzh2023}. At the same time, in all these implementations, the algorithm was embedded in the devices. When we discuss the implementation of the technique  for ``universal'' quantum devices such as IBMQ quantum computers or similar, it is important to minimize the number of quantum gates in the circuit implementation of the automaton with respect to the architecture restrictions.  Many types of quantum computers (for example, quantum devices based on superconductors) do not allow us to apply two-qubit gates to an arbitrary pair of qubits. They have a specific architecture of qubit connectivity that is represented by a qubit connectivity graph. Vertices of the graph correspond to qubits, and two-qubit gates can be applied only to qubits corresponding to vertices connected by an edge. 
The basic implementation part of quantum hashing circuits is a uniformly controlled rotation (UCR) operator. The CNOT cost (the number of CNOT gates) of this operator with respect to a qubit connectivity graph was previously discussed in \cite{mottonen2005decompositions,bvms2005mottonen2006decompositions} for the linear nearest-neighbor (LNN) architecture and in \cite{zkk2023} for several examples of more complex graphs. Note that the UCR operator itself is useful in different areas such us universal gate decomposition \cite{mottonen2005decompositions,bvms2005mottonen2006decompositions}, circuit for machine learning models \cite{ksz2025,skzk2023} and others.

K\={a}lis \cite{kalis18} suggested a shallow circuit for the automaton with 3 qubits recognizing the $MOD_{11}$ language. Later, in a general way, the approach was developed by Ziatdinov et. al. \cite{ziiatdinov2023gaps,zkk2025}. It allows us to reduce the CNOT cost. Computational experiments show that we can find the required parameters of the hash function with the same number of qubits as for the standard circuit for the quantum hashing algorithm.  Therefore, the CNOT cost of the shallow circuit can be exponentially less than for the standard circuit. At the same time, the theoretical exponential superiority of the shallow circuit is not shown \cite{ziiatdinov2023gaps,zkk2025}. By the way, the method is very promising for current and near-future quantum devices that definitely cannot support a huge number of CNOT gates. The efficient implementation of a shallow circuit for an automaton that recognizes the $MOD_p$ language was proposed by Khadieva et al. \cite{ksy2024} for devices based on the LNN architecture, by Khadieva \cite{k2024aliya} for a more complex architecture, that is, a cycle with tails (like a ``sun'' and  ``two joint suns'') represented by 16-qubit and 27-qubit Eagle r3 IBMQ architectures. Khadiev et al. \cite{kkcw2025} suggested a method to construct a circuit for an arbitrary connected graph.
Vasiliev \cite{v2023} discussed a similar method but for Rz gates instead of Ry gates which efficient was shown in computational experements \cite{kmak2025}.

In this paper, we present a general method that allows us to develop a quantum circuit for the quantum hashing algorithm for an arbitrary qubit connectivity graph. The CNOT cost of the constructed circuit is between $2n\ell -4\ell +2$ and $6n\ell -7\ell+2$ depending on the complexity of the graph, where $n$ is the number of qubits and $\ell$ is the length of the input word.
It is better than the existing technique \cite{kkcw2025} that provides a circuit with the CNOT cost from $3n\ell-4\ell+2$ to $6n\ell-4\ell+2$ depending on the complexity of the graph. Additionally, we show a simple example of the qubit connectivity graph (a grid graph that is a popular qubit connectivity graph \cite{googlewillow2025,gao2025Zuchongzhi,rigetti}) such that our algorithm gives a better result compared to the result of \cite{kkcw2025}. When we apply our approach to the LNN architecture, we obtain a circuit with the same CNOT cost as the circuit specially built for the LNN architecture \cite{ksy2024}. The same situation we have if we consider the case of 16-qubit and 27-qubit Eagle r3 IBMQ architectures \cite{k2024aliya}.
Our algorithm uses a solution of the shortest non-simple covering path problem, which is a modification of the shortest covering path problem \cite{cpr1994}. We present exact and approximate solutions for the problem that can be independently interesting to a reader.
 We consider the method with applications for  automata (data stream processing algorithms) for two unary languages $MOD_p=\{a^\ell: \ell \bmod p=0\}$ and the $EQ_p=\{a^\ell b^r: \ell=r \pmod p\}$ equality language that are simple enough but has all the required properties to demonstrate the quantum hashing (fingerprinting) algorithm.
%Our method allows us to construct a circuit with .... CNOT cost. 

The structure of this paper is the following.
Section \ref{sec:prelims} describes the required notations and preliminaries.
% Section \ref{sec:ucr} provides an algorithm for constructing a quantum circuit for the uniformly controlled rotation gate that is a base for the standard circuit implementation of quantum hashing and particularly the automaton for the $MOD_p$ language.
Section \ref{sec:shallow} provides an algorithm for generating the shallow quantum circuit for quantum hashing. An application of the method for automata is presented in Section \ref{sec:qfamodp}.
 Graph theory tools are presented in Section \ref{sec:tools}.
The final Section \ref{sec:concl} concludes the paper and contains some open questions.

%%%%%%%%%%%%%%%%%%%%%%%%%%%%%%%%%%%%%%%%
%%    Preliminaries
%%%%%%%%%%%%%%%%%%%%%%%%%%%%%%%%%%%%%%%%
\section{Preliminaries}\label{sec:prelims}
%\subsection{Graph Theory}
\textbf{Graph Theory.}
Let us consider an undirected unweighted graph $G=(V,E)$, where $V$ is the set of vertices and $E$ is the set of undirected edges. Let $n=|V|$ be the number of vertices, and $m=|E|$ be the number of edges. 

A non-simple path $P$  is a sequence of vertices $(v_{i_1},\dots,v_{i_h})$ that are connected by edges, that is $(v_{i_j},v_{i_{j+1}})\in E$ for all $j\in\{1,\dots,h-1\}$. Note that a non-simple path can contain duplicates.
Let the length of the path be the number of edges in the path $len(P)=h-1$.
A path $P=(v_{i_1},\dots,v_{i_h})$ is called simple if there are no duplicates among $v_{i_1},\dots,v_{i_h}$. 
The distance $dist(v,u)$ is the length of the shortest path between vertices $v$ and $u$. %Typically, when we say just a ``path'', we mean a ``simple path''. 
Let $\textsc{Neighbors}(v)$ be a list of neighbors for a vertex $v$, i.e., $\textsc{Neighbors}(v)=(u_{i_1},\dots,u_{i_k})$ such that $(v,u_{i_j})\in E$, and $|\textsc{Neighbors}(v)|=k$ be the length of the list.

%Let us consider an undirected weighted graph $S=(V',E')$, where $V'$ is the set of vertices, and $E'$ is the set of undirected weighted edges. Let $w:V'\times V'\to \mathbb{R}$ be a weight function. We assume that $w(v,u)=\infty$ if $(v,u)\not\in E$.

%For a path $P=(v_{i_1},\dots,v_{i_h})$, the length  $w(P)$ is the sum of edge weights in the path, i.e. $w(P)=\sum\limits_j^{h-1}w(v_{i_j},v_{i_{j+1}})$.
%%%%%%%%%%%%%%%%%%%%%%%%%%
\textbf{The Shortest Non-simple 1-covering Path Problem.}
Let us consider an undirected unweighted connected graph $G=(V,E)$ such that $n=|V|$ is a number of vertices and $m=|E|$ is a number of edges.

Let us consider the ``Shortest non-simple 1-covering path'' problem that is a modification of the well-known ``Shortest covering path'' problem (SCPP or SCP problem) \cite{cpr1994}.
The ``Shortest non-simple 1-covering path'' problem (1-SNSCPP or 1-SNSCP problem) is defined as follows.

 Let $P=(v_{i_1},\dots,v_{i_k})$ be a \textbf{non-simple}  path. We say that the path covers all visiting vertices and vertices that are connected with visited vertices by one edge. Formally, the path $P$ covers a set of vertices $R_P$ such that any vertex $v$ from this set is either
%\begin{itemize}
    %\item 
    (i) $v$ belongs to $P$ (there is $j\in\{1,\dots,k\}$ such that $v=v_{i_j}$);
    %\item 
    (ii) $v$ is connected with a vertex from $P$ (there is $j\in\{1,\dots,k\}$ such that $(v,v_{i_j})\in E$).
%\end{itemize}
Let $B_P=R_P\backslash\{v_{i_1},\dots,v_{i_k}\}$, i.e. they are vertices connected with visited vertices by one edge. 
If the path $P$ covers all vertices ($R_P=V$), then we call it a 1-covering path. The solution of the 1-SNSCP problem is the shortest 1-covering path.

As the SCP problem, the 1-SNSCP problem has a strong connection with the Hamiltonian path problem and the Travelling salesman problem \cite{cormen2001}. Any connected graph has a 1-covering path.
The decision version of the SCP problem is NP-complete \cite{cpr1994}. The Travelling salesman problem (TSP) is NP-hard. Similarly, by polynomial reduction of TSP to 1-SNSCPP, we can show that it is NP-hard.  
%
%\begin{theorem}\label{th:whp-np-cmpl}
%The (3,2,1)-CP problem is NP-hard. % (See Appendix \ref{apx:whp-np-cmpl})
%\end{theorem}
%
Let us estimate the maximum possible length of a 1-covering path.

\begin{lemma}\label{lm:len-wnsh}
The length of a 1-covering path in a connected graph $G$ of $n$ vertices is at most $2n-3$.  (See Appendix \ref{apx:len-wnsh}).
\end{lemma}
The algorithm for the problem is presented in Section \ref{sec:tools}.

%%%%%%%%%%%%%%%%%%%%%%%%%
%\subsection{Quantum circuits}\label{sec:qcirc}
\textbf{Quantum circuits.}
Quantum circuits consist of qubits and a sequence of gates applied to these qubits. A state of a qubit is a column-vector from ${\cal H}^2$ Hilbert space. It can be represented by $a_0|0\rangle+a_1|1\rangle$, where $a_0,a_1$ are complex numbers such that $|a_0|^2+|a_1|^2=1$, and $|0\rangle$ and $|1\rangle$ are unit vectors. Here we use the Dirac notation. A state of $n$ qubits is represented by a column-vector from ${\cal H}^{2^n}$ Hilbert space. It can be represented by $\sum_{i=0}^{2^n-1}a_i|i\rangle$, where $a_i$ is a complex number such that $\sum_{i=0}^{2^n-1}|a_i|^2=1$, and $|0\rangle,\dots |2^n-1\rangle$ are unit vectors. Graphically, on a circuit, qubits are presented as parallel lines. 

As basic gates, we consider the following ones:

  $H=\frac{1}{\sqrt{2}}\begin{pmatrix}
1 & 1 \\
1 & -1 
\end{pmatrix}$,     $X=\begin{pmatrix}
0 & 1 \\
1 & 0 
\end{pmatrix}$,
 $R_y(\xi)=\begin{pmatrix}
\cos(\xi/2) & -\sin(\xi/2) \\
\sin(\xi/2) & \cos(\xi/2) 
\end{pmatrix}$,

$R_z(\xi)=\begin{pmatrix}
e^{\frac{i\xi}{2}} & 0 \\
0 & e^{-\frac{i\xi}{2}} 
\end{pmatrix}$,
$CNOT=\begin{pmatrix}
1 & 0 & 0 & 0\\
0 & 1 & 0 & 0\\
0 & 0 & 0 & 1\\
0 & 0 & 1 & 0 
\end{pmatrix}$.  

Additionally, we consider four non-basic gates. The first three gates are

\noindent
$CR_y(\xi)=\begin{pmatrix}
1 & 0 & 0 & 0\\
0 & 1 & 0 & 0\\
0 & 0 & \cos(\xi/2) & -\sin(\xi/2) \\
0 & 0 & \sin(\xi/2) & \cos(\xi/2) 
\end{pmatrix}$,
$CR_z(\xi)=\begin{pmatrix}
1 & 0 & 0 & 0\\
0 & 1 & 0 & 0\\
0 & 0 & e^{\frac{i\xi}{2}}  & 0\\
0 & 0 & 0 & e^{-\frac{i\xi}{2}} 
\end{pmatrix}$, 

$SWAP=\begin{pmatrix}
1 & 0 & 0 & 0\\
0 & 0 & 1 & 0\\
0 & 1 & 0 & 0\\
0 & 0 & 0 & 1 
\end{pmatrix}$.

The fourth is a uniformly controlled rotation gate on $n-1$ control qubits and one target qubit. We denote it by $UCR^{n-1}_b$ gate. Here, $b$ can be $y$ or $z$. Let $|\psi\rangle$ be an $(n-1)$-qubit quantum control register, and $|\phi\rangle$ be one target qubit. The gate applies the $R_a(\xi_i)$ rotation gate to $|\phi\rangle$ if the control qubit register is $|i\rangle$ for $i\in\{0,\dots, 2^{n-1}-1\}$. If $a=y$, then the gate applies $R_y(\xi_i)$.  If $a=z$, then the gate applies $R_z(\xi_i)$.
%
%$CR_y(\xi)=\begin{pmatrix}
%1 & 0 & 0 & 0\\
%0 & 1 & 0 & 0\\
%0 & 0 & cos(\xi/2) & -sin(\xi/2) \\
%0 & 0 & sin(\xi/2) & cos(\xi/2) 
%\end{pmatrix}$.

The reader can find more information about quantum circuits in \cite{nc2010,aazksw2019part1,k2022lecturenotes}.

%%%%%%%%%%%%%%%%%%%%%%%%%%%%%%%%%%%%%%%%%%%%%%%%%%
% Method for Constructing a Circuit for Quantum Hashing
%%%%%%%%%%%%%%%%%%%%%%%%%%%%%%%%%%%%%%%%%%%%%%%%%%
\section{Method for Constructing a Circuit for Quantum Hashing}\label{sec:hash}

In this section, we present a method that allows us to construct a circuit for the quantum fingerprinting or quantum hashing algorithm for a connected graph $G=(V,E)$ that is a qubit connectivity graph for a device. 
Two simple examples of application of the quantum hashing algorithm are the quantum automata for the unary $MOD_p=\{a^\ell: \ell \bmod p=0$, $p$ is prime$\}$ language and for the unary Equality language $EQ_p=\{a^\ell b^r: \ell=r \pmod p\}$. That are presented in Section \ref{sec:qfamodp}. More information about the quantum hashing (quantum fingerprinting) algorithm can be found in Section \ref{sec:qfamodp} and Appendix \ref{apx:hash}.

The $n$-qubit quantum circuit for the quantum hashing algorithm contains three main parts. We assume that we have qubits $q_0,\dots,q_{n-1}$.
The first part is to apply the Hadamard transformation $H$ to qubits with indexes $1,\dots, n-1$. The second part is to apply the main $U_a$ transformation of the quantum hashing algorithm. Often, the main transformation is applied several times (let it be $\ell$ times). In the case of quantum automaton for $MOD_p$, for the input word $a^\ell$, we apply the transformation $\ell$ times\footnote{In the case of  $EQ_p$ language, we apply it $\ell+r$ times.}. The final part is to apply the Hadamard transformation $H$ to qubits with indexes $1,\dots, n-1$, and to measure all qubits.
The standard implementation of the $U_a$ transformation is the $UCR^{n-1}_b$ gate. At the same time, in terms of minimization of the CNOT cost, the shallow circuit is more efficient. It is discussed in Section \ref{sec:shallow}.

\subsection{The Quantum Finite Automata for $MOD_p$ and $EQ_p$ Languages}\label{sec:qfamodp}

The general description of the quantum hashing or quantum fingerprinting technique from \cite{af98,an2008,an2009,akv2008,bcwd2001} can be found in Appendix \ref{apx:hash}.  This technique is well-known and allows us to compute a short hash or fingerprint that identifies a (potentially long) string of data with high probability.
Here we present a quantum finite automaton (data stream processing algorithm) for the unary $MOD_p=\{a^\ell: \ell$ mod $p=0$, $p$ is prime$\}$ language as a simple example of this technique. At the same time, it shows the power and has all the main properties of the technique. 
A quantum finite automaton (QFA) for a unary language is a tuple $(Q, Q_{acc}, q_0, U_a, U_{\$}, U_{\cent})$. Here $Q$ is a set of $2^n$ quantum basis states ($n$ qubits), $Q_{acc}\subset Q$ is the set of accepting states, $q_0\in Q$ is a starting state; $U_a, U_{\$}$ and $ U_{\cent}$ are unitary transformations on $2^n$ states.
The automaton processes a $\$ \omega \cent$ word for an input word $\omega=a^\ell$ for some positive integer $\ell$, $\$$ is the  left end marker and $\cent$ is the right end marker. The automaton works with $n$-qubit register, starts from the $|q_0\rangle$ state. We apply $U_{\$}$ to the register on the left end marker, then apply $U_{a}$ on each symbol of $\omega$, and finally apply  $U_{\cent}$ on the right end marker. Finally, we measure the quantum register and accept the word if the result state belongs to $Q_{acc}$. The probability of this event is $\sum_{q_i\in Q_{acc}}|\alpha_i|^2$, where $\alpha_i$ is the amplitude of the $i$-th state. A language is recognizable by an automaton if the automaton accepts member words with probability $1-\varepsilon$, and rejects non-members with probability $1-\varepsilon$ for some $0<\varepsilon<0.5$. 

The automaton for $MOD_p$ language \cite{an2008,an2009} is such that $n=\log_2((2/\varepsilon) \ln 2p)$, $|q_0\rangle=|0\rangle$, $Q_{acc}=\{|0\rangle\}$.
The transformations $U_{\$}$ and $U_{\cent}$ are Hadamard transformations $H$ on each qubit except the $0$-th one. The matrix $U_a$ is the $UCR_y^{n-1}$ gate with angles $\xi_i=\frac{2\pi k_i}{p}$ for $k_0,\dots, k_{2^{n-1}-1}\in\{1,\dots,p-1\}$. Here $k_i$ are good coefficients from \cite{an2008} discussed in Appendix \ref{apx:hash}.
Note that  $p$ can be any integer. At the same time, for the prime $p$, quantum automata can use exponentially less memory comparing to their classical counterparts \cite{af98}. There are results \cite{AY12,agky16,av2009,kkk2022,kk2017,av2011,av2013} that use fingerprinting for non-prime $p$. 

The automaton for $EQ_p$ language is a modification of the automaton for $MOD_p$ language \cite{agky16,av2009,kkk2022,kk2017,av2011,av2013}. We also have $n=\log_2((2/\varepsilon) \ln 2p)$, $|q_0\rangle=|0\rangle$, $Q_{acc}=\{|0\rangle\}$.
The transformations $U_{\$}$ and $U_{\cent}$ are Hadamard transformations $H$ on each qubit except the $0$-th one. The matrix $U_a$ is the $UCR_y^{n-1}$ gate with angles $\xi_i=\frac{2\pi k_i}{p}$ for $k_0,\dots, k_{2^{n-1}-1}\in\{1,\dots,p-1\}$.
The matrix $U_b$ is the $UCR_y^{n-1}$ gate with angles $\xi'_i=-\xi_i$.
So, in fact, we compute $\ell-r$ and check whether $\ell-r=0$ $($mod $p)$ which is exactly what the automaton for $MOD_p$ does.
At the same time, the $EQ_p$ language shows us the ``equality checking ability'' of the quantum hashing technique.

%%%%%%%%%%%%%%%%%%%%%%%%%%%%%%%%%%%%%%%%%%55
\subsection{Shallow Circuit for Quantum Hashing.}\label{sec:shallow}
\begin{figure}[H]
\includegraphics[width=0.4\textwidth]{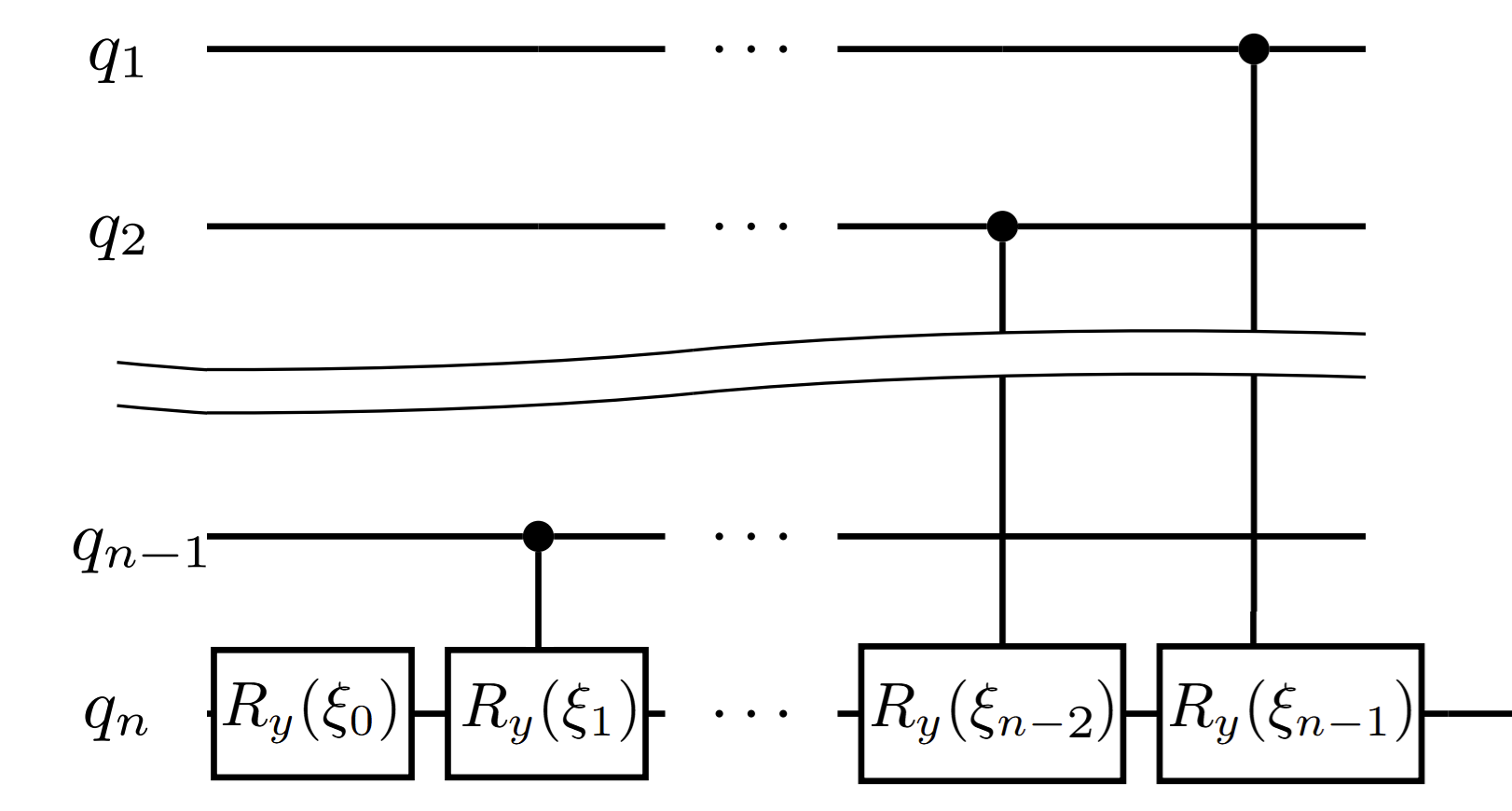}% Here is how to import EPS art
\caption{\label{fig:qf} Shallow circuit for quantum fingerprinting or quantum hashing algorithm}
\end{figure}

Here we consider a shallow circuit \cite{kalis18,ziiatdinov2023gaps,zkk2025} for $n-1$ control qubits. If we do not have restrictions for applying two-qubit gates (when $G$ is a complete graph or a ``star'' graph), then the shallow circuit for the main transformation of the quantum hashing algorithm is present in Figure \ref{fig:qf}. In the case of the shallow circuit, we call the transformation $U_s$. Here we assume that we have qubits $q_0,\dots,q_{n-1}$ such that $q_1,\dots,q_{n-1}$ are control ones and $q_{0}$ is the target one.
We can consider this transformation as an approximation for the $UCR_y^{n-1}$ gate.

Here, we assume that we have a solution for the 1-SNSCP problem presented in Section \ref{sec:tools}.
Let $q_i$ be the logical qubits of the original circuit. Let the qubits of a physical device be associated with the vertices of the graph $G$, and we call them $v_i$.
The algorithm for constructing a circuit is the following.

%\begin{itemize}
   % \item[] 
     \noindent
   \textbf{Step 1.} We find the shortest 1-covering path in the graph $G$ using the algorithm from Section \ref{sec:tools}. Assume that the path is $P=(v_{i_1}, \dots,v_{i_k})$.
   
    %\item[] 
    \noindent
    \textbf{Step 2.} The target qubit $q_{0}$ corresponds to the vertex $v_{i_1}$. The control qubits $q_1,\dots,q_{n-1}$ correspond to other vertices.  We assume that we have a set ${\cal U}$ of control qubits that are already used. Initially, it is empty ${\cal U}\gets\emptyset$. Let $j$ be the index of the element in the path $P$ that corresponds to the target qubit. Initially, $j\gets 1$.
    
    % \item[] 
    In the next steps, the target qubit travels along the path $P$. 
    
    %\item[]
      \noindent
    \textbf{Step 3.} We apply the control rotation with the control $v'$ and the target $v_{i_j}$ qubits, where $v'\in \textsc{Neighbors}(v_{i_j})\backslash\{v_{i_{j+1}}\}$ and $v'\not\in {\cal U}$. In other words, $v'$ is a neighbor of $v_{i_j}$ but not $v_{i_{j+1}}$, and $v'$ is not visited. Then, we add $v'$ to the set ${\cal U}$, i.e. ${\cal U}\gets {\cal U}\cup\{v'\}$.
   
    %\item[] 
      \noindent
    \textbf{Step 4.} If $v_{i_{j+1}}\not\in {\cal U}$, then we apply the control rotation to the control $v_{i_{j+1}}$ and the target $v_{i_j}$ qubits. After that, we add $v_{i_{j+1}}$ to the set ${\cal U}$, i.e. ${\cal U}\gets {\cal U}\cup\{v_{i_{j+1}}\}$. 
    
    % \item[]
      \noindent
    \textbf{Step 5.} We apply the SWAP gate to $v_{i_{j}}$ and $v_{i_{j+1}}$. After that, we update $j\gets j+1$ because the value of the target qubit moves to $v_{i_{j+1}}$. If $j<k-1$, then we go 
    to Step 3 and go to Step 6 otherwise.
    
     %\item[] 
       \noindent
     \textbf{Step 6.} We do this step if $j=k$. We apply the control rotation with the control $v'$ and the target $v_{i_k}$ qubits, where $v'\in \textsc{Neighbors}(v_{i_k})$ and $v'\not\in {\cal U}$. Then, we add $v'$ to the set $U$, i.e. ${\cal U}\gets {\cal U}\cup\{v'\}$. This is the final step of the algorithm.  
%\end{itemize}

The implementation of the algorithm is presented in Appendix \ref{apx:algo-impl}. Let us discuss the CNOT cost of the generated circuit.

Let us discuss a circuit for $\ell$ applications of the main quantum hashing transformation $U_s$ (as an example, it is the main part of the circuit implementation of the automaton for $MOD_p$ processing $a^\ell$). For an odd application of $U_s$ we apply the presented algorithm. For an even application of $U_s$, we do the same algorithm, but move the target qubit in the reverse order from $v_k$ to $v_1$. Note that the vertices $v'\in \textsc{Neighbors}(v_{i_k})$ are also considered in the reverse order. That is why, we have two sequential $\textsc{cR}(u,v_{i_k})$ gates, where  $\textsc{cR}(u,v_{i_k})$ represents the $CR_y(\xi)$ gate with $u$ as a control qubit, $v_{i_k}$ as the target qubit, and $\xi$ is the angle corresponding to $u$. Here $u$ is the last vertex in the list $\textsc{Neighbors}(v_{i_k})$. So, we can merge these two operators and use only one with the double angle. We have a similar situation with the first element from $\textsc{Neighbors}(v_{i_1})$.

\begin{figure}[H]
\includegraphics[width=0.4\textwidth]{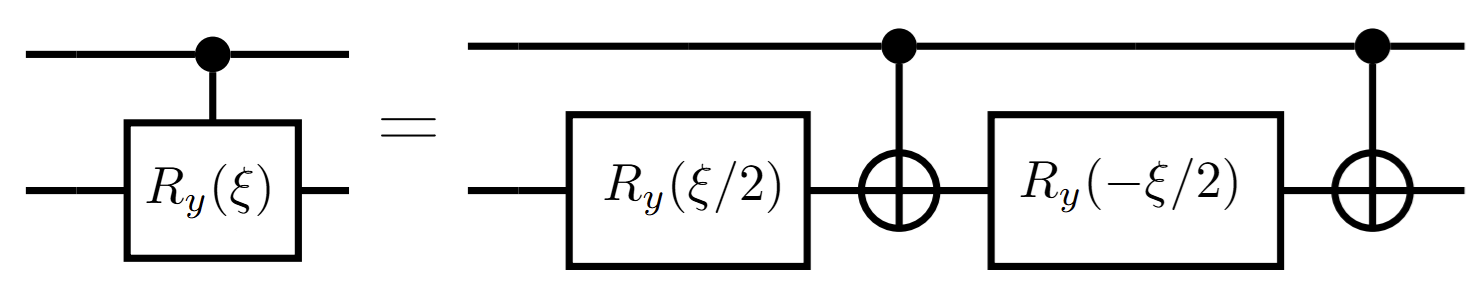}% Here is how to import EPS art
$\quad$
\includegraphics[width=0.27\textwidth]{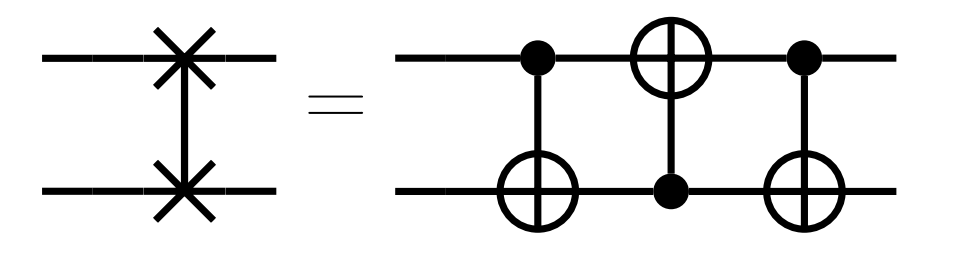}
%$\quad$
\caption{\label{fig:cr} Representation of $CR_y$ and  $SWAP$ gates using only basic gates}
\end{figure}

We can represent the control rotation operator $\textsc{cR}(u,v)$ and the corresponding angle $\xi$ as a sequence of $\textsc{R}(v,\xi/2)$, $\textsc{cnot}(u,v)$, $\textsc{R}(v,-\xi/2)$, and $\textsc{cnot}(u,v)$, (see Figure \ref{fig:cr}, the left circuit). Here $\textsc{R}(v,\xi/2)$ is the $R_y(\xi/2)$ gate applied to the qubit $v$; $\textsc{cnot}(u,v)$ is the $CNOT$ gate for $u$ as a control and $v$ as a target. The $\textsc{swap}(u,v)$ gate can be represented as a sequence $\textsc{cnot}(u,v)$, $\textsc{cnot}(v,u)$, and $\textsc{cnot}(u,v)$ (see Figure \ref{fig:cr}, the right circuit). 

\vspace{-0.5cm}

\begin{figure}[H]
\includegraphics[width=0.4\textwidth]{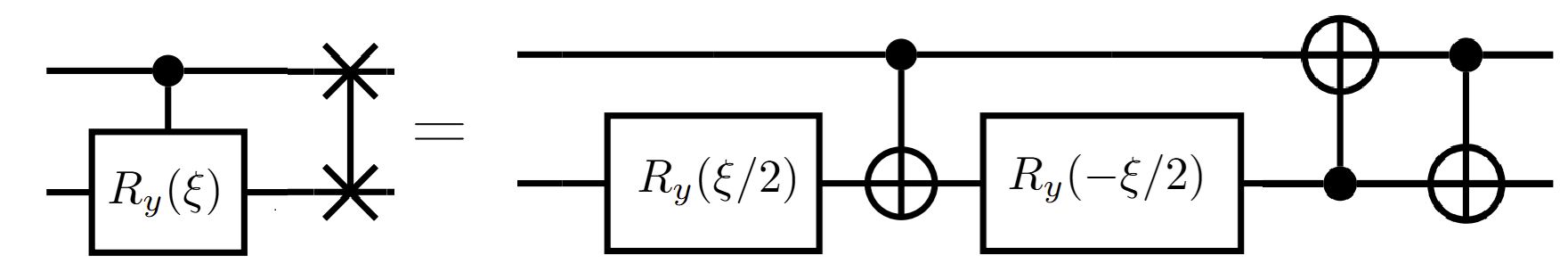}% Here is how to import EPS art
\caption{\label{fig:crswap1} Representation of a pair $CR_y$ and $SWAP$ gates using only basic gates}
\end{figure}

\vspace{-0.2cm}
Two sequential operators $CR_y$ and $SWAP$ that are $\textsc{cR}(u,v)$ and $\textsc{swap}(u,v)$ procedures can be represented by a circuit in Figure \ref{fig:crswap1}. Note that two sequential $\textsc{cnot}(u,v)$ gates are annihilated.

\vspace{-0.5cm}

\vspace{-0.2cm}
Let us look at the CNOT cost of these operators, that is, the number of CNOT gates in the circuit.
We can say that CNOT cost of $\textsc{cR}(u,v)$ is $2$; CNOT cost of $\textsc{swap}(u,v)$ is $3$; CNOT cost of two sequential operators $\textsc{cR}(u,v)$ and  $\textsc{cnot}(u,v)$ is $3$. 
Finally, we can discuss the CNOT cost of the constructed circuit for $\ell$ applications of quantum hashing operator $U_s$.
\begin{theorem}\label{th:qh1}
    The CNOT cost of the circuit for $\ell$ applications of the quantum hashing operator $U_s$ generated by the presented algorithm is $(3k + 2(n-k'))\ell-5\ell + 2$, where $k$ is the length of the 1-covering path $P=(v_{i_1},\dots,v_{i_k})$, and $k'$ is the number of distinct vertices in $P$, and $n$ is the number of vertices in the qubit connectivity graph. 
    (See Appendix \ref{apx:qh1})
\end{theorem}

Let us estimate the minimal and maximal possible CNOT cost.

\begin{corollary}\label{cr:path}
The CNOT cost of the circuit for $\ell$ applications of the quantum hashing operator $U_s$ generated by the presented algorithm is between $2n\ell-4\ell+2$ and $6n\ell-7\ell+2$, where $k$ is the length of the 1-covering path $P=(v_{i_1},\dots,v_{i_k})$, and $k'$ is the number of distinct vertices in $P$. 
(See Appendix \ref{apx:path})
\end{corollary}

\subsubsection{Comparing with Existing Results}\label{sec:compare-qh}
\begin{figure}[H]
\begin{center}    \includegraphics[width=0.7\textwidth]{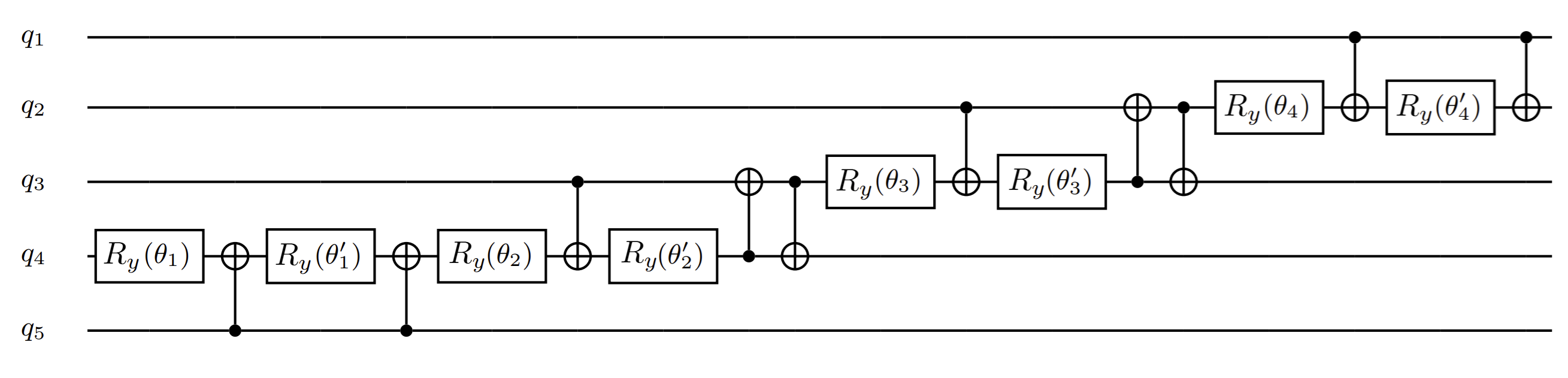}
\end{center}
% Here is how to import EPS art
\caption{\label{fig:lnn-qhash} A quantum circuit for one application of operator $U_s$ for Quantum hashing (quantum fingerprinting) algorithm in the case of $5$ qubits LNN architecture device}
\end{figure}

Let us compare our generic approach with existing circuits for specific graphs. The LNN architecture is a common layout graph for many quantum devices. The graph is a chain where $v_1$ is connected to $v_2$;  $v_i$ is connected to $v_{i-1}$ and $v_{i+1}$ for $i\in\{2,\dots,n-1\}$; $v_n$ is connected with $v_{n-1}$.
The quantum fingerprinting (quantum hashing) algorithm for the LNN architecture was discussed in \cite{ksy2024}. The best 1-covering path is $P=(v_2,\dots,v_{n-1})$.
So, our method gives us the same result circuit as in  \cite{ksy2024}. The circuit for the quantum fingerprinting algorithm on $5$ qubits is presented in Figure \ref{fig:lnn-qhash}. The CNOT cost of the circuit is presented in the Lemma \ref{lm:lnn-cnotcost}.

\vspace{-0.5cm}

\vspace{-0.2cm}
\begin{lemma}\label{lm:lnn-cnotcost}
 The CNOT cost of the circuit for $\ell$ applications of the quantum hashing operator $U_s$ generated by the presented algorithm in the case of LNN architecture is $3n\ell-7\ell+2$.
\end{lemma}

\begin{figure}[H]
\begin{center}
\includegraphics[width=0.2\textwidth]{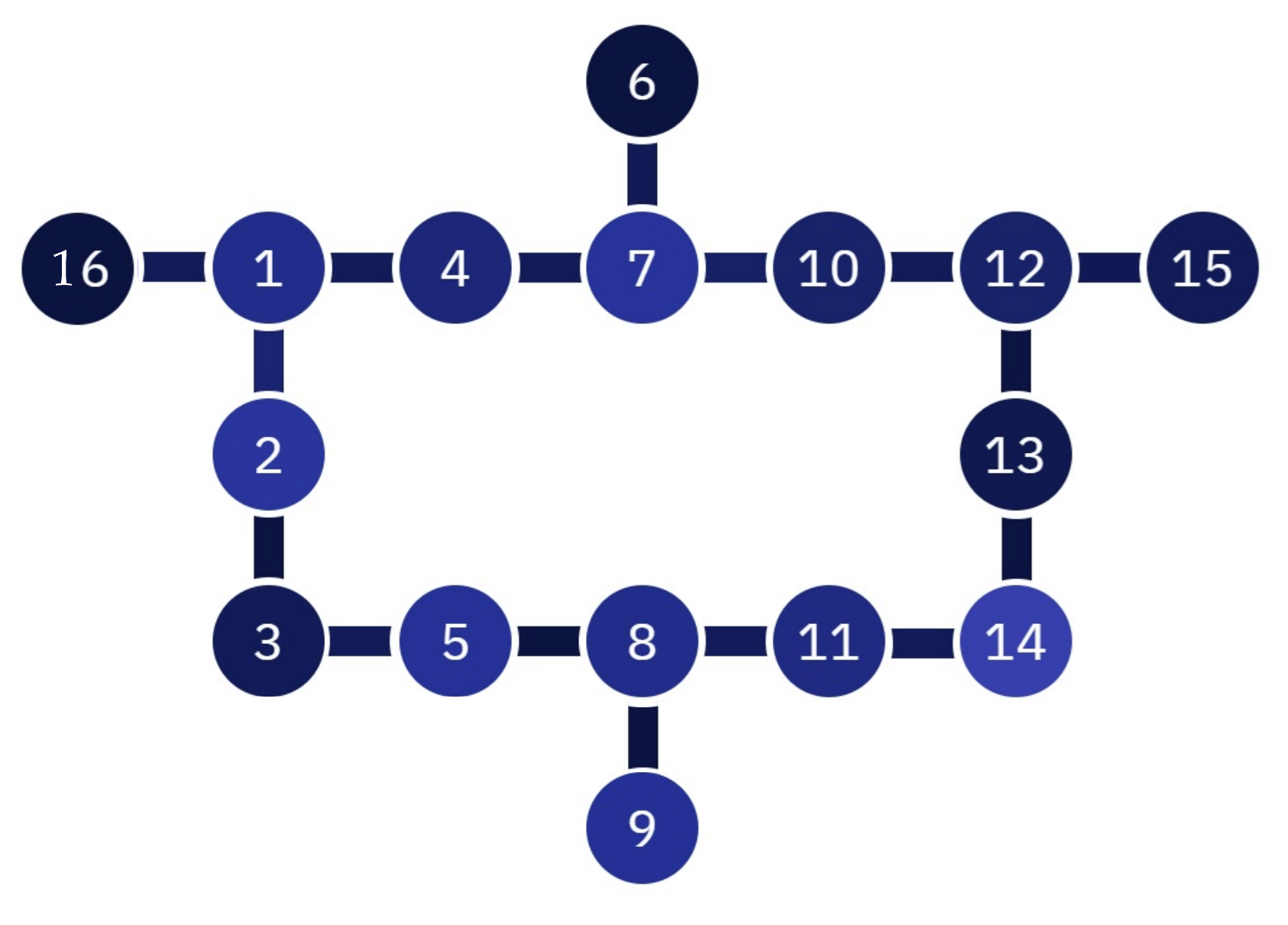}% Here is how to import EPS art
$\quad$
\includegraphics[width=0.37\textwidth]{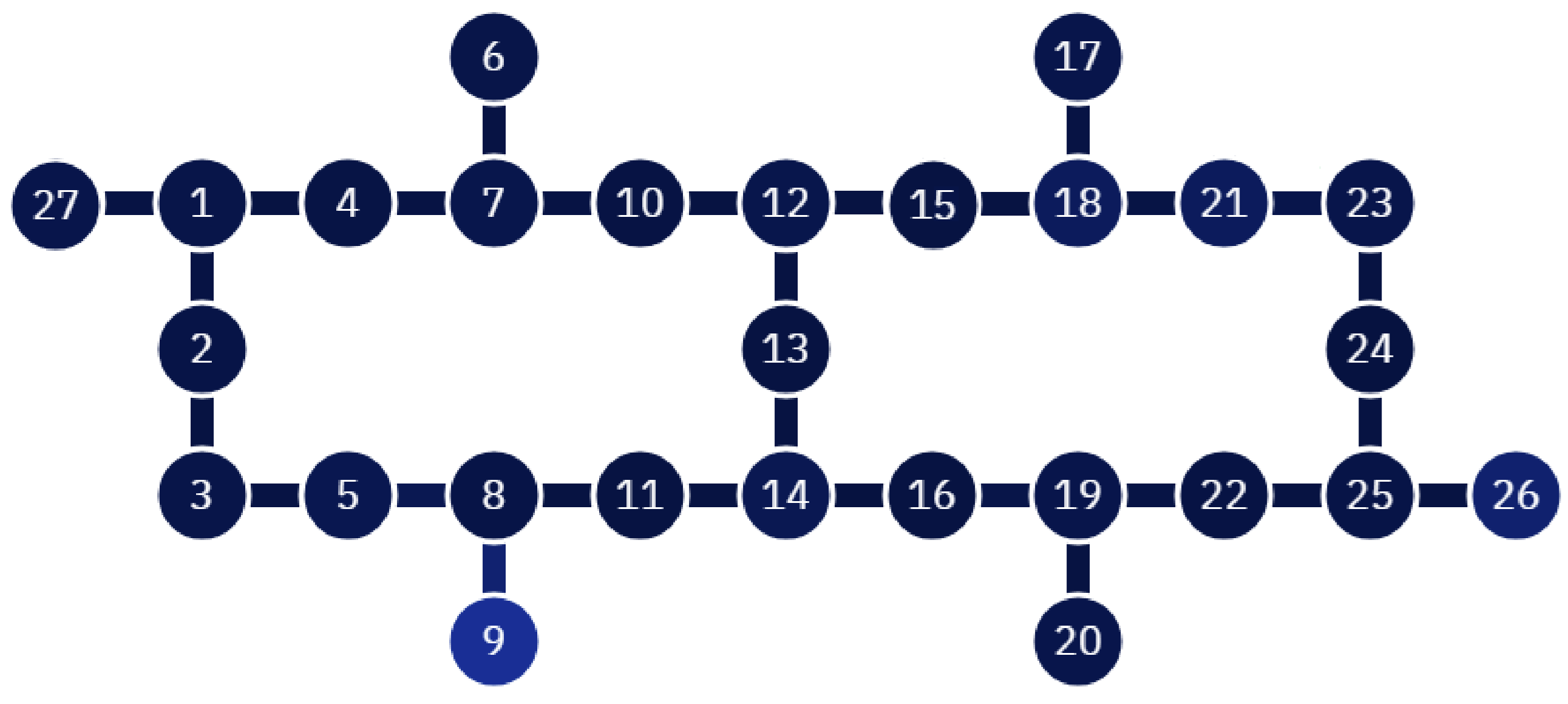}
\end{center}
\caption{\label{fig:sun1} 
16-qubit Falcon r4P and 27-qubit Falcon r5.11 architectures of IBMQ}
\end{figure}

Let us consider more complex graphs like a cycle with tails (like a ``sun'' or ``two joint suns''). The 16-qubit Falcon r4P and 27-qubit Falcon r5.11 IBM machines of these architecture were considered in \cite{k2024aliya}. The graphs are
presented in Figure \ref{fig:sun1}. Our generic method gives the same circuits as the circuits specially constructed for these devices \cite{k2024aliya}. 

\vspace{-0.5cm}

\begin{figure}[H]
\begin{center}
\includegraphics[width=0.25\textwidth]{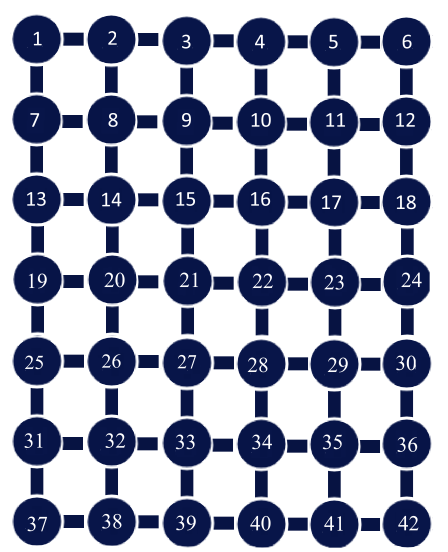}% Here is how to import EPS art
$\quad$
\includegraphics[width=0.25\textwidth]{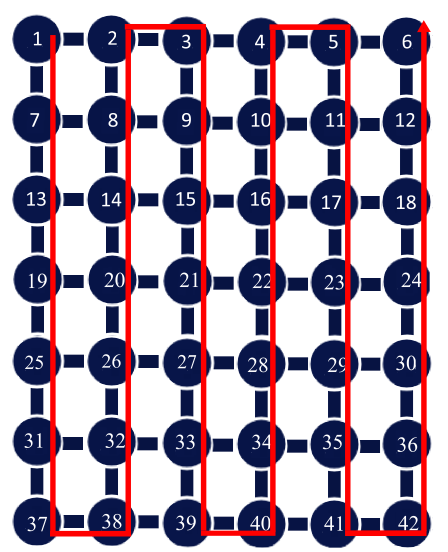}
$\quad$
\includegraphics[width=0.25\textwidth]{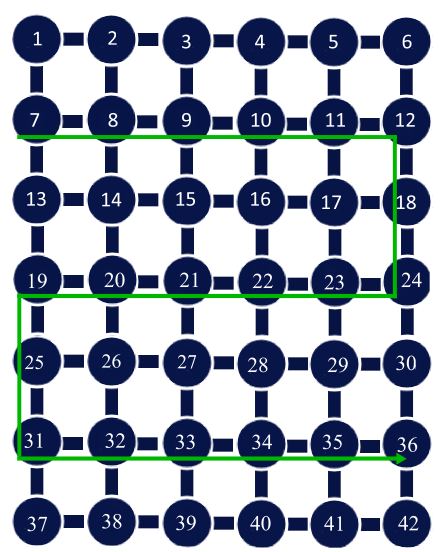}
\end{center}
\caption{\label{fig:grid} 
A grid architecture with $7$ rows and $6$ columns. The green line is the shortest 1-covering path (for our algorithm). The red line is the shortest path that visits all vertices at least once (for algorithm of \cite{kkcw2025}).}
\end{figure}

\vspace{-0.5cm}
When we compare our result with \cite{kkcw2025}, we can see that the maximal possible CNOT cost of a circuit can be $6n\ell-7\ell+2$ for our method and $6n\ell-4\ell+2$ for \cite{kkcw2025}. Let us present an example of a graph in which our approach generates a circuit much better than the algorithm of \cite{kkcw2025}. Let us consider a grid with $q$ rows and $t$ columns (an example of a grid with 3 rows and 6 columns is presented in Figure \ref{fig:grid}) such that $q$ is odd, and $t\geq 3$. For simplicity of analysis, we assume that $t$ is an even integer. 

\vspace{-0.5cm}
Our algorithm constructs a path that visits $(t+1)\frac{q-1}{2}-1$ vertices of the even rows. So, each edge of the path gives 3 CNOT gates, and each other vertex gives us 2 CNOT gates. Let $\ell=1$.
We get $3((t+1)\frac{q-1}{2}-2)$ CNOT gates because of the path, and $2\cdot (qt-((t+1)\frac{q-1}{2}-1))$ CNOT gates because of the rest vertices. The total CNOT coast is $3((t+1)\frac{q-1}{2}-2) + 2\cdot (qt-((t+1)\frac{q-1}{2}-1)) = 2qt + \frac{(t+1)(q-1)}{2}-3 = 2qt+0.5qt +0.5(q-1-t)  -3=2.5qt+0.5(q-1-t) - 3$.

The algorithm of \cite{kkcw2025} constructs a path that visits all vertices at least once. The length of the path is $qt-1$. So, the CNOT cost of the circuit is $3qt-3$.  So, the difference is $3qt-3 - (2.5qt+0.5(q-1-t) - 3)=0.5qt + 0.5(t+1-q)$. That is more than $16\%$ difference.

In the case of $q=7$ and $t=6$, our algorithm gives a circuit with $101$ CNOT gates, and the algorithm of \cite{kkcw2025} gives a circuit with $123$ CNOT gates.  
%Example of the graph where our approach is better is gric 3xt.
%Our gives (t-1)*3+2*t= 5t-3.
%Old gives (t-2+1+t-1+1+t-2)*3+4=6t-2
%\nocite{*}
%Example of the graph where our approach is better is gric 3xt.
%Our gives (t-1)*3+2*t= 5t-3.
%Old gives (t-2+1+t-1+1+t-2)*3+4=6t-2
 
%%%%%%%%%%%%%%%%%%%%%%%%%%%%%%%%%%%%%%%%%%%%%%%%%%%
%                    Tools
%%%%%%%%%%%%%%%%%%%%%%%%%%%%%%%%%%%%%%%%%%%%%%%%%%%5

\section{Algorithm for the Shortest Non-simple 1-covering Path Problem}\label{sec:tools}
Let us have an unweighted undirected graph $G=(V,E)$, where $n=|V|$, and $m=|E|$.
%Secondly, we 
Let us provide the algorithm for the 1-SNSCP problem.
First, let us present a procedure $\textsc{ShortestPaths}(G)$ that constructs two $n\times n$-matrices $W$ and $A$ by a graph $G$. 
Elements of the matrix $W$ are lengths of the shortest paths between each pair of vertices in $G$, i.e. $W[v,u]=dist(v,u)$.
The matrix $A$ represents the shortest paths between the vertices of $G$. The element $A[v,u]$ is the last vertex in the shortest path between $v$ and $u$; and $A[v,v]=NULL$. In other words, if $t=A[v,u]$, then $P_{v,u}=P_{v,t}\circ u$, where $P_{v,u}$ is the shortest path between $v$ and $u$. Based on this fact, we can present a procedure $\textsc{GetShortestPath}(v,u)$ that computes $P_{v,u}$ using the matrix $A$. Note that the implementation does not add the first element of the path $P_{v,u}$ because we do not need it in our algorithm. The implementation of the procedure is presented in Appendix \ref{apx:getpath}.
%Algorithm \ref{alg:getpath}. %(See Appendix \ref{apx:getpath})

%\vspace{-0.5cm}

%Let the procedure $\textsc{GetPathNoFirst}(v,u)$ returns $\tilde{P}_{v,u}$ that is the path $P_{v,u}$ without the first element. The implementation is the same, but without the $P_{v,u}\gets (v)\circ P_{v,u}$ line.

%\vspace{-0.5cm}
We can construct these two matrices using $n$ invocations of the Breadth First Search (BFS) algorithm \cite{cormen2001}. The total time complexity for construction of the matrices is $O(n^3)$. The algorithm for constructing $A$ and $W$ is presented in Appendix \ref{apx:floyd} for completeness of presentation.

Let us define a function $D:2^{V}\times V\to\{0,\dots,n,\infty\}$ such that $D(S,v)$ is the length of the shortest path $P$ that visits all the vertices of $S$ and the last vertex is $v$. Formally, $P=(v_{i_1},\dots,v_{i_k})$, $ v_{i_k}=v$, $S\subset\{v_{i_1},\dots,v_{i_k}\}$. If there is no such path, then $D(S,v)=\infty$. Note that the path $P$ is non-simple, and it can visit vertices from $V\backslash S$.
Let us present an algorithm for computing $D(S,v)$ for each $S\in2^V$ and $v\in S$. It is easy to see that $D(\{v\},v)=0$ for each $v\in V$. For other pairs $(S,v)$, we compute it using the following statement $D(S,v)=\min\{D(S\backslash\{v\},u)+W[u,v]: u\in S\}$. 
To construct the path itself, we define a function $F:2^{V}\times V\to V\cup\{NULL\}$ such that $F(S,v)$ is the vertex that precedes $v$ in the path that visits all vertices of $S$. Formally,  $F(S,v)=\min\{i:D(S\backslash\{v\},v_i)+W[v_i,v]=D(S,v), (v_i,v)\in E\}$. If there is no such vertex $v_i$, then $F(S,v)=NULL$.
So, we can compute $F(S,v)$ together with $D(S,v)$, $F(S,v)=u$ if $u=argmin\{D(S\backslash\{v\},u)+W[u,v]: u\in S\}$. If $D(S,v)=\infty$, then $F(S,v)=NULL$.
This idea allows us to define a recursive procedure $\textsc{ComputeNSD}(G,v)$ whose implementation is presented in Appendix \ref{apx:computeNSD}.
%
%Algorithm \ref{alg:computeNSD}. %(See Appendix \ref{apx:computeNSD}).
%
Let us present the procedure $\textsc{GetNSPath}(S,v)$ that returns the path that visits all vertices of $S$ and ends in $v$. The procedure collects the path using $\textsc{GetShortestPath}$ between the vertices obtained from $F$.
The implementation of $\textsc{GetNSPath}(S,v)$ is presented in 
Appendix \ref{apx:getpath2}.
%Algorithm \ref{alg:getpath2}.% (See Appendix \ref{apx:getpath2}).

%\vspace{-0.5cm}

%\vspace{-0.5cm}
Furthermore, we define a function $C:2^V\to\{0,1\}$ such that $C(S)=1$ iff $V=S\cup\{v:v\in V\backslash S, $ and there is $u\in S$ such that $(u,v)\in E\}$. In other words, $C(S)=1$ if all vertices of $V\backslash S$ are connected to vertices of $S$ by one edge. Let us define a procedure $\textsc{ComputeC}(G)$ that computes the function $C$. For this reason, we compute a set $R=S\cup \bigcup_{v\in S}\{u: u\in \textsc{Neighbors}(v)\}$. We do it for each set $S\in2^V$. The implementation of the procedure is presented in Algorithm \ref{alg:checkWHP}. (See Appendix \ref{apx:checkWHP}).

Now we are ready to define the whole algorithm for the 1-SNSCP problem. Firstly, we form the functions $D$, and $F$.  We choose the shortest $P$, where $P=\textsc{GetNSPath}(S,v)$ is a path that visits all the vertices of $S$ at least once and $C(S)=1$, which means that $P$ is a 1-covering path. Note that $P$ can visit not only the vertices of $S$. That is why we choose the largest $S$ for the shortest path $P$. It visits only vertices from $S$ in that case.

Let $\textsc{Shortest1CP}(G)$ be the procedure that returns the shortest 1-covering path. The implementation of the procedure is presented in Appendix \ref{apx:wnshpath}.%Algorithm \ref{alg:wnshpath}.  %Appendix \ref{apx:wnshpath}.
The correctness and complexity of the algorithm is discussed in Theorem \ref{th:wnshpath}.

%\vspace{-0.5cm}

%\vspace{-0.5cm}
\begin{theorem}\label{th:wnshpath}
    The presented algorithm solves the 1-SNSCP problem, and the time complexity is $O((m+n)2^n)$. (See Appendix \ref{apx:twnshpath})
\end{theorem}
 
%%%%%%%%%%%%%%%%%%%%%%%%%%%%%%%%%%%%%%%%%%
% Approximate Algorithm for 1-SNSCP Problem
%%%%%%%%%%%%%%%%%%%%%%%%%%%%%%%%%%%%%%%%

\subsection{Approximate Algorithm for the Shortest Non-simple 1-covering Path Problem}

We are planning to use the solution of the problem for optimization of a circuit for the quantum hashing algorithm. So for big $n$, the current solution is too slow.

Due to the strong connection of the 1-SNSCP problem with the Travelling salesman problem (TSP) and The Shortest Covering Path problem (SCPP), we can use heuristic algorithms, for example, Ant colony optimization \cite{dg1997}, or greedy algorithms like \cite{jm1997} that are used for the TSP or algorithms used for SCPP \cite{cpr1994}. 

Here we present a fast approximate solution for the problem that can be used for practical applications.
Let us define two subtasks:

%\begin{itemize}
    %\item 
    (i) The Connected Dominating Set problem (CDS problem). For a given graph $G=(V,E)$, we want to find a connected set $S$ of minimal size such that $V=S\cup B$, where $B=\{u: u\in \textsc{Neighbors}(v)$ for some $v\in S\}$. Informally, each vertex of the graph either belongs to $S$ or connected to a vertex from $S$ by one edge.
    
    %\item 
    (ii) For a given weighed graph $G'=(V',G')$, the shortest non-simple path that visits all vertices of the graph at least once.
%\end{itemize}

The first problem can be solved using a $(\ln \Delta + 3)$-approximating algorithm from \cite{gk1998}, where $ \Delta = max\{|\textsc{Neighbors}(v)|: v\in V\}$ is the maximal number of neighbors of a vertex from $V$. Here, $\alpha$-approximating algorithm means the result is at most $\alpha$ times bigger than the solution. Properties of the algorithm are described in the next lemma.
\begin{lemma}[\cite{gk1998}]\label{lm:cds}
    There is an $(\ln \Delta + 3)$-approximate algorithm for CDS problem. Time complexity of the algorithm is $O((n+m)\log n)$.
\end{lemma}
 
The second problem can be solved by an analog of the Christofides–Serdyukov algorithm \cite{c2022,s1978,vs2020}. Let us consider a spanning tree of the graph $G=(V,E)$. It is a tree $T=(V,E')$, where $E'\subset E$. %The covering path in $G$ should pass all vertices. So, w
We can construct a non-simple path $P$ that is the Euler tour \cite{cormen2001} of the tree $T$. The path visits all the vertices of the graph $G$, but possibly it is not the shortest. The length of the path is $2|V|-2$. The length of the minimal possible path that visits all vertices is at least $|V|-1$. So, the algorithm gives us at most $2$ times longer path. The solution is $2$-approximating solution for the second problem.

\begin{lemma}\label{lm:tcsns}
    The time complexity of the presented $2$-approximate algorithm for searching the shortest non-simple path that visits all vertices of the graph at least once is $O(|V|+|E|)$.
\end{lemma}
\Beginproof
    The spanning tree can be constructed using the depth-first search algorithm with $O(|V|+|E|)$ time complexity \cite{cormen2001}. The Euler tour \cite{cormen2001} can also be done with $O(|V|+|E|)$ time complexity.
\Endproof

So, the whole algorithm has two steps: 

%\begin{itemize}
    %\item[] 
    \textbf{Step 1.} Construction the smallest connected domain $S$ of the graph $G$. Then, consider the subgraph $G(S)=(S,E(S))$, where $E(S)\subset E$ is the set of edges of $G$ that connects only vertices from $S$. We use the $(\ln \Delta + 3)$-approximate algorithm from Lemma \ref{lm:cds}.
    %\item[]
    
    \textbf{Step 2.} We construct a path that visits all vertices at least once in the graph $G(S)$.  We use the $2$-approximate algorithm from Lemma
    \ref{lm:tcsns}.
%\end{itemize}

We claim that the presented algorithm solves the 1-SNSCP problem and it is a  $2(ln \Delta + 3)$-approximate algorithm.

\begin{theorem}\label{th:aprox}
    The presented algorithm solves 1-SNSCP problem, it is a $2(\ln \Delta + 3)$-approximate algorithm, and the time complexity is $O((n+m)\log n)$.
    (See Appendix \ref{apx:aprox})
\end{theorem}
Note that the maximal number of neighbors $\Delta$ in current devices is often small (it can be $2,3,4$ or $5$ if we consider IBM or Rigetti quantum devices). That is why $\ln \Delta$ can be at most $2$. Additionally, if we use the approximate solution to the  problem, then the length of the path can be longer, but it cannot be longer than $2n$ according to Lemma \ref{lm:len-wnsh}.

\section{Conclusion}\label{sec:concl}
In the paper we presented an algorithm for generating a shallow quantum circuit for quantum hashing algorithm with respect to minimization of the CNOT cost. We present exact algorithm with exponential time complexity and approximate algorithm with almost linear time complexity. We show that our algorithm generates better quantum circuit comparing to existence approaches \cite{kkcw2025}. As an example we consider a grid graph and obtain about $16\%-17\%$  improvement. As a base of the generating algorithm we use the 1-SNSCP problem.

We have several open questions regarding to the obtained results:

%\begin{enumerate}
    %\item 
    (i)Can we present an exact solution of the 1-SNSCP problem with polynomial time complexity for special classes of graphs? As an example of graphs it can be qubit connectivity graphs used by IMBQ or Rigetti quantum computers.
    
    (ii) Can we present an efficient greedy solutions for the 1-SNSCP problem?
%\end{enumerate}
 \bibliographystyle{unsrt}
 \bibliography{tcs}

 \newpage
\appendix

%%%%%%%%%%%%%%%%%%%
\section{Quantum Fingerprinting or Quantum Hashing}\label{apx:hash}
Let us present some basic concepts of quantum fingerprinting technique from \cite{af98,an2008,an2009,akv2008,bcwd2001}. This technique is well-known and allows us to compute a short hash or fingerprint that identifies a (potentially long) string of data with high probability.

For the problem being solved we choose some positive integer $m$, an error probability bound $\varepsilon > 0$, fix $t = \lceil(2/\varepsilon) \ln 2m\rceil$, and construct a mapping $g : \{0, 1\}^n\to \mathbb{Z}$. Then for arbitrary binary string $\sigma = (\sigma_1 \dots \sigma_n)$ we create it's fingerprint $|h_\sigma\rangle$ composing $t$ single qubit fingerprints $|h_\sigma^i\rangle$:
\[|h_\sigma^i\rangle=cos\frac{2\pi k_i g(\sigma)}{m}|0\rangle + sin\frac{2\pi k_i g(\sigma)}{m}|1\rangle,\]\[
|h_\sigma\rangle=\frac{1}{\sqrt{t}}\sum_{i=1}^{t}|i\rangle|h^i_{\sigma}\rangle\]

Here the last qubit is rotated by $t$ different angles about the $\hat{y}$ axis of the Bloch sphere. The chosen parameters $k_i \in\{1\dots,m-1\}$, for $i\in\{1\dots t\}$ are ``good'' in the following sense. A set of parameters $K = \{k_1,\dots, k_t\}$ is called ``good'' for $g\neq 0 \mod m$ if
\[\frac{1}{t^2}\left(\sum_{i=1}^t cos\frac{2\pi k_i g}{m}\right)^2<\varepsilon\]
The left side of the inequality is the squared amplitude of the basis state $|0\rangle^{\otimes \log_2 t} |0\rangle$ if the operator
$H^{\otimes \log_2 t}\otimes I $ has been applied to the fingerprint $|h_\sigma\rangle$. Informally, that kind of set guarantees, that
the probability of error will be bounded by a constant below $1$.

The following lemma from \cite{akv2008,an2008,an2009} proves the existence of a ``good'' set.
\begin{lemma}[\cite{akv2008}]
There is a set $K$ with $|K| = t = \lceil(2/\varepsilon) \ln 2m\rceil$  which is ``good'' for all $g\neq 0 \mod m$.
\end{lemma}

We use this result for fingerprinting technique \cite{akv2008} choosing the set $K = \{k_1,\dots, k_t\}$ that is ``good'' for all $g = g(\sigma)\neq 0$. It allows us to distinguish those inputs whose image is $0$ modulo $m$ from the others.

That hints at how this technique may be applied:
\begin{enumerate}
\item We construct $g(x)$, that maps all acceptable inputs to $0$ modulo $m$ and others to arbitrary non-zero (modulo $m$) integers.

\item After the necessary manipulations with the fingerprint, the $H^{\otimes \log_2 t}$ operator is applied to the first $\log_2 t$ qubits. This operation ``collects'' all cosine amplitudes at the all-zero state. That is, we obtain the state of the type
\[|h'_\sigma\rangle=\frac{1}{t}\sum_{i=1}^{t}cos\left(\frac{2\pi k_i g(\sigma)}{m}\right) |00\dots 0\rangle|0\rangle + \sum_{i=2}^{2t}\alpha_i|i\rangle\]
\item This state is measured on the standard computational basis. Then we accept the input if the outcome is the all-zero state. This happens with the probability
\[Pr_{accept}(\sigma)=\frac{1}{t^2} \left(\sum_{i=1}^{t}cos\frac{2\pi k_i g(\sigma)}{m}\right)^2,\]
which is $1$ for the inputs, whose image is $0 \mod m$ and is bounded by $\varepsilon$ for the others.
\end{enumerate}

Due to \cite{ziiatdinov2023gaps,zkk2025,kalis18}, the algorithm can be implemented using the shallow circuit presented in Figure \ref{fig:qf}. In that case, the angles $\frac{2\pi k_i}{m}$ should be linear combinations of $\xi_j$. Due to \cite{ziiatdinov2023gaps,zkk2025}, it is not known whether we can keep the same number of qubits or should we increase the number of qubits exponentially. At the same time, computational experiments show \cite{ziiatdinov2023gaps,zkk2025} that we can find enough good parameters $\xi_i$ such that $t$ qubits are enough for $\varepsilon$ error probability.
%%%%%%%%%%%%%%%%
\section{Proof of Lemma \ref{lm:len-wnsh}.}\label{apx:len-wnsh}
\textbf{Lemma 
 \ref{lm:len-wnsh}} {\em
The length of a 1-covering path in a connected graph $G$ of $n$ vertices is at most $2n-3$. 
}

\Beginproof
Let us consider a spanning tree of the graph $G=(V,E)$. It is a tree $T=(V,E')$, where $E'\subset E$. %The covering path in $G$ should pass all vertices. So, w
We can construct a non-simple path $P$ that is the Euler tour \cite{cormen2001} of the tree $T$ but does not visit the leaves of the tree. The path covers all the vertices of the graph $G$, but maybe be it is not the shortest. Each edge (except edges incident to leaves)  in the tour is visited at most twice (in the up and down direction). Therefore, the length of the path $len(P)\leq 2n-{\cal L}$, where ${\cal L}$ is the number of leaves, and ${\cal L}\geq 2$. So, we obtain the bound for the number of vertices in the path $2n-2$, and for the length of the path, the bound is $2n-3$.
\Endproof

\section{Implementation of the Procedure $\textsc{ShortestPaths}$ for Shortest Paths Searching}\label{apx:floyd}
Here we discuss how to construct matrices $W$ and $A$ such that $W[v,u]$ is the length of the shortest path between vertices $v$ and $u$, and $A[v,u]$ is the last vertex in the shortest path between $v$ and $u$; and $A[v,v]=NULL$. The procedures are simple, but we present them for the completeness of the results representation.

Firstly, we present a procedure $\textsc{SingleSrcShortestPath}(v)$ that finds the shortest paths for a single source vertex $v$ that is based on the BFS algorithm \cite{cormen2001}. The algorithm calculates the $v$-th rows of $W$ and $A$. The implementation is presented in Algorithm \ref{alg:bfs1}. Here we assume that we have a queue data structure \cite{cormen2001} that allows us to do the next actions in constant time:
\begin{itemize}
    \item $\textsc{Add}(queue, v)$ adds an element to the queue;
    \item $\textsc{Remove}(queue)$ removes an element from the queue and returns the element;
    \item  $\textsc{Init}()$ returns an empty queue;
     \item $\textsc{isEmpty}(queue)$ returns $True$ if the queue is empty and $False$ otherwise.
\end{itemize}
\begin{algorithm}[H]
\caption{Implementation of $\textsc{SingleSrcShortestPath}(v)$}\label{alg:bfs1}
\begin{algorithmic}
\State $queue\gets \textsc{Init}()$
\State  $\textsc{Add}(queue, v)$
\For{$u\in V$}
\State $W[v,u]\gets\infty$
\State $A[v,u]\gets NULL$
\EndFor
\State $W[v,v]\gets 0$
\While{$\textsc{isEmpty}(queue)=False$}
\State $t\gets\textsc{Remove}(queue)$
\For{$r\in \textsc{Neighbors}(t)$}
\If{$W[v,r]=\infty$}
    \State $A[v,r]\gets t$
    \State $W[v,r]=W[v,t]+1$
    \State $\textsc{Add}(queue, r)$
\EndIf
\EndFor
\EndWhile
\end{algorithmic}
\end{algorithm}
As an implementation of the $\textsc{ShortestPaths}$ procedure, we invoke $\textsc{SingleSrcShortestPath}(v)$ for each vertex $v\in V$.
\begin{algorithm}[H]
\caption{Implementation of $\textsc{ShortestPaths}(G)$ for a $G=(V,E)$ graph}\label{alg:bfs2}
\begin{algorithmic}
\State 
\For{$v\in V$}
    \State $\textsc{SingleSrcShortestPath}(v)$
\EndFor
\State \Return $(W,A)$
\end{algorithmic}
\end{algorithm}
\begin{lemma}
    Time complexity of the $\textsc{ShortestPaths}$ procedure is $O(n^3)$.
\end{lemma} 
\Beginproof
Time complexity of BFS is $O(n+m)=O(n^2)$ due to \cite{cormen2001}.  Invocation of $n$ BFS algorithms for each $v\in V$ is $O(n^3)$. 
\Endproof

%%%%%%%%%%%%%%

\section{Implementation of $\textsc{ComputeC}(G)$}\label{apx:checkWHP}

\begin{algorithm}[H]
\caption{Implementation of $\textsc{ComputeC}(G)$}\label{alg:checkWHP}
\begin{algorithmic}
\For{$S\in 2^V$} 
\State $R=S$
\For{$v\in S$}
\For{$u\in \textsc{Neighbors}(v)$}
\State $R\gets R\cup\{u\}$
\EndFor
\EndFor
\If{$|R|=n$}
\State $C(S)\gets 1$
\Else
\State $C(S)\gets 0$
\EndIf
\EndFor
\end{algorithmic}
\end{algorithm}

%%%%%%%%%%%%%%%%%%%%%%%%%%%%%

\section{Proof of Theorem \ref{th:wnshpath}}\label{apx:twnshpath}
\textbf{Theorem \ref{th:wnshpath}. }{\em
    The presented algorithm solves the 1-SNSCP problem, and the time complexity is $O((m+n)2^n)$. 
}

\Beginproof
%\Beginproof
    Let us show the correctness of the algorithm. Suppose that the algorithm finds the shortest path $P$ that visits all vertices of $S$ such that $C(S)=1$, and $S$ is the largest for this length of $P$. Assume that there is 1-covering path $P'=(v_{i_1},\dots,v_{i_{k'}})$ that is shorter than $P$. Let $S'=\{v_{i_1},\dots,v_{i_{k'}}\}$, then $\textsc{GetNSPath}(S',v_{i_{k'}})=P'$. It means $len(P')=D(S',v_{i_{k'}})<len(P)$ because $P$ is the shortest path among all paths computed by $\textsc{GetNSPath}(S,v)$. This claim contradicts the assumption $len(P)>len(P')$.

    The procedure $\textsc{ComputeNSD}$ is invoked once for each subset $S\in 2^V$ and vertex $v\in V$. The time complexity of all invocations of the procedure is $O((m+n)\cdot 2^n)$.  
    The time complexity of the \textsc{ShortestPaths} procedure is $O(n^3)$.
    The time complexity for the procedure $\textsc{ComputeC}$ is $O((m+n)2^n)$ because we check all subsets $S\in 2^V$ and check at most $m$ edges of the graph for each subset.
    
    The complexity of $\textsc{GetNSPath}$ is $O(n)$ because the maximal length of the path is $2n$ due to Lemma \ref{lm:len-wnsh}.
    So, the total complexity is $O(n^3+(m+n)\cdot 2^n+(m+n)\cdot 2^n+n)=O((m+n)2^n)$.
\Endproof

%%%%%%%%%%%%%%%%%%%%%%%%%%%%%%

\section{Implementation of the Algorithm for Constructing a Quantum Circuit}\label{apx:algo-impl}
Let us present a procedure that implements the algorithm described in Section \ref{sec:hash}. We assume that we have $\textsc{cR}(u,v)$ procedure that applies the control rotation operator $CR_y$ to $u$ as a control qubit and $v$ as a target one. The angle $\xi_r$ corresponds to the qubit $q_r$ associated with the vertex $v$. Additionally, we have $\textsc{swap}(u,v)$ procedure that applies swap gate to $u$ and $v$ qubits.

\begin{algorithm}[H]
\caption{Implementation of $\textsc{ConstructForPath}(P)$ procedure. Algorithm of constructing circuit for quantum hashing or quantum fingerprinting for a path $P=(v_{i_1}, \dots,v_{i_k})$}\label{alg:cascade}
\begin{algorithmic}
\State $P=(v_{i_1}, \dots,v_{i_k})\gets \textsc{Shortest1CP}(G)$\Comment{Step 1}
\State ${\cal U}\gets\emptyset$\Comment{Step 2}
\For{$j\in\{1,\dots,k-1\}$}
\For{$v'\in\textsc{Neighbors}(v_{i_j})$}\Comment{Step 3}
\If{$v'\not\in {\cal U}$ and $v'\neq v_{i_{j+1}}$}
\State  $\textsc{cR}(v',v_{i_j})$
\State ${\cal U}\gets {\cal U}\cup\{v'\}$
\EndIf
\EndFor
\If{$v_{i_{j+1}}\not\in {\cal U}$}
\State  $\textsc{cR}
(v_{i_{j+1}},v_{i_j})$\Comment{Step 4}
\State ${\cal U}\gets {\cal U}\cup\{v_{i_{j+1}}\}$
\EndIf
\State $\textsc{swap}(v_{j},v_{j+1})$\Comment{Step 5}
\EndFor
\For{$v'\in\textsc{Neighbors}(v_{i_k})$}\Comment{Step 6}
\If{$v'\not\in {\cal U}$}
\State  $\textsc{cR}(v',v_{i_k})$
\State ${\cal U}\gets {\cal U}\cup\{v'\}$
\EndIf
\EndFor
\end{algorithmic}
\end{algorithm}

\section{The Proof of the Theorem \ref{th:aprox}}\label{apx:aprox}

\textbf{Theorem \ref{th:aprox}} {\em
    The presented algorithm solves 1-SNSCP problem, it is a $2(ln \Delta + 3)$-approximate algorithm, and the time complexity is $O((n+m)\log n)$.
}

\Beginproof
Let us consider the solution $P=(v_{i_1},\dots,v_{i_k})$ for 1-SNSCP problem for some graph $G=(V,E)$. The set $S=\{v_{i_1},\dots,v_{i_k}\}$ is the set of vertices visited by $P$. Note that all vertices of the graph are either belongs to $V$ or connected to a vertex from $S$ with one edge. Let $S_d$ be the solution of CDS problem for the graph. Therefore, the size $|S|\geq |S_d|$.
The length of the path $P$ is $len(P)=\geq |S|\geq |S_d|$.

Let us consider the solution obtained by the approximate solution for the 1-SNSCP problem.
Let $S'_d$ be the approximate solution of the first part (for CDS problem). So, $|S'_d|\leq (ln\Delta +3)|S_d|$.

Let the path $P'$ be the approximate solution of the second part (the shortest non-simple path that visits all vertices of $S'_d$ at least once). The length of the path is $len(P')\leq 2|S'_d|\leq 2(ln\Delta +3)|S_d|$.
So, we can say, that $len(P')\leq 2(ln\Delta +2)|S_d|$, and $len(P)\geq |S_d|$. Therefore, $len(P')\leq len(P)\cdot 2(ln\Delta +2)$.

The time complexity of the solution is $O((n+m)\log n)$ for the first part, and $O(|S_d'|+E(S_d'))=O(n+m)$ for the second part. The total time complexity is $O((n+m)\log n)$.
\Endproof

\section{The Proof of the Theorem \ref{th:qh1}}\label{apx:qh1}

\textbf{Theorem \ref{th:qh1}} {\em
    The CNOT cost of the circuit for $\ell$ applications of quantum hashing operator $U_s$ generated by presented algorithm is $(3k + 2(n-k'))\ell-5\ell + 2$, where $k$ is the length of the 1-covering path $P=(v_{i_1},\dots,v_{i_k})$, and $k'$ is the number of distinct vertices in $P$, and $n$ is the number of vertices in the qubit connectivity graph. 
}

\Beginproof
If we look at the algorithm, then we can see that it is a sequence of one of three options:
%\begin{itemize}
    %\item 
    (i) a pair of $CR_y$ and $SWAP$ gates if $v_{j+1}\not\in U$;
    %\item
    (ii) $SWAP$ gate if $v_{j+1}\in U$;
    %\item 
    (iii) $CR_y$ gate for  $v'$ and $v_{i_{j+1}}$.
%\end{itemize}
%
The CNOT cost for pairs of $CR_y$ and $SWAP$ gates is $3$; for the $SWAP$ gate, it is $3$; for the $CR_y$ gate, it is $2$.
On applying $U_s$ once, the number of steps in which we apply pairs of gates $CR_y$ and SWAP or the SWAP gate is $k-1$ because we move the target qubit by the path of length $k$. We apply a single  $CR_y$ (without SWAP gate) to vertices that do not belong to $P$. The number of such nodes is $n-k'$.
At the same time, for two sequential applications of the $U_s$ operator, one $CR_y$ operator disappears, as we discussed before.
So, the CNOT cost of the first $\ell-1$ operators $U_s$ is $(3(k-1) + 2(n-k'-1))(\ell - 1)$, and the CNOT cost of the last operator is $3(k-1) + 2(n-k')$. The total cost is $(3k + 2(n-k'))\ell-5\ell + 2$.
\Endproof

\section{The Proof of the Corollary \ref{cr:path}}\label{apx:path}

\textbf{Corollary \ref{cr:path}} {\em
The CNOT cost of the circuit for $\ell$ applications of the quantum hashing operator $U_s$ generated by the presented algorithm is between $2n\ell-4\ell+2$ and $6n\ell-7\ell+2$, where $k$ is the length of the 1-covering path $P=(v_{i_1},\dots,v_{i_k})$, and $k'$ is the number of distinct vertices in $P$. 
}

\Beginproof
The minimum possible CNOT cost is in the case when the length of $P$ is minimal. If we have a vertex that is connected with all other vertices of the graph, then we can construct $P$ of length $1$. So, $k=1$ and $k'=1$. Therefore, the total CNOT cost is $2n\ell-4\ell+2$.

Due to Lemma \ref{lm:len-wnsh}, the maximum length of the path is $k=2n-2$. The maximum possible number of distinct vertex in $P$ is $n-2$. So, the total CNOT cost is $(3(2n-2)+4)\ell-5\ell+2=6n\ell-7\ell+2$.
\Endproof

\section{Implementation of the Procedure $\textsc{GetShortestPath}$ for Geting the Shortest Path}\label{apx:getpath}

\begin{algorithm}[H]
\caption{Implementation of $\textsc{GetShortestPath}(v,u)$}\label{alg:getpath}
\begin{algorithmic}
\State $t\gets A[v,u]$
\If{$v=u$}
\State $P_{v,u}\gets()$\Comment{an empty list}
\Else
\State $P_{v,u}\gets (u)$
\While{$t\neq v$}
\State $P_{v,u}\gets t\circ P_{v,u} $
\State $t\gets  A[v,t]$
\EndWhile
\EndIf
%\State $P_{v,u}\gets (v)\circ P_{v,u}$
\State \Return $P_{v,u}$
\end{algorithmic}
\end{algorithm}

\section{Implementation of the Procedure $\textsc{ComputeNSD}$ for Computing $D$ and $F$}\label{apx:computeNSD}
\begin{algorithm}[H]
\caption{Implementation of $\textsc{ComputeNSD}(S,v)$}\label{alg:computeNSD}
\begin{algorithmic}
\If{$S=\{v\}$}
        \State $D(S,v)\gets 0$, $F(S,v)\gets NULL$
\Else 
    \State $D(S,v)\gets \infty$, $F(S,v)\gets NULL$
    \For{$u\in S$}
            \If{$D(S\backslash\{v\},u)$ is not computed}
            \State $\textsc{ComputeNSD}(S\backslash\{v\},u)$
            \EndIf
            \If{$D(S\backslash\{v\},u)+W[u,v]<D(S,v)$}
                \State $D(S,v)\gets D(S\backslash\{v\},u)+W[u,v]$, $F(S,v)\gets u$
            \EndIf        
    \EndFor
\EndIf
\end{algorithmic}
\end{algorithm}

\section{Implementation of the Procedure $\textsc{GetNSPath}$ for Collecting a Path using $F$}\label{apx:getpath2}
\begin{algorithm}[H]
\caption{Implementation of $\textsc{GetNSPath}(S,v)$}\label{alg:getpath2}
\begin{algorithmic}
\State $P=()$ \Comment{We initialize it by an empty list}
\While{$F(S,v)\neq NULL$}
\State $u\gets F(S,v)$, $S\gets S\backslash\{v\}$
\State $P\gets \textsc{GetShortestPath}(u,v)\circ P$\Comment{We add $P_{u,v}$ path without the vertex $u$ to the begin of the list}
\State $v\gets u$
\EndWhile
\State $P\gets v\circ P$
\State \Return $P$
\end{algorithmic}
\end{algorithm}

\section{Implementation of the Procedure $\textsc{Shortest1CP}$ for solving the 1-SNSCP Problem}\label{apx:wnshpath}
\begin{algorithm}[H]
\caption{Implementation of $\textsc{Shortest1CP}(G)$}\label{alg:wnshpath}
\begin{algorithmic}
\State $\textsc{ShortestPaths}(G)$
\State $\textsc{ComputeC}(G)$
\State $S'\gets \emptyset, v'\gets NULL, \ell\gets \infty$ 
\For{$S\in 2^V$}
\For{$v\in S$}
\State $\textsc{ComputeNSD}(S,v)$

\If{$C(S)=1$}
\If{$\ell>D(S,v)$ or ($\ell=D(S,v)$ and $|S|>|S'|$) }
\State $\ell\gets D(S,v), S'\gets S, v'\gets v$
\EndIf
\EndIf
\EndFor
\EndFor
\State $P\gets \textsc{GetNSPath}(S',v')$
\State \Return $P$
\end{algorithmic}
\end{algorithm}
\end{document}